\documentclass[prd,aps,floats,preprint,preprintnumbers]{revtex4}
\usepackage{amsmath,amssymb}
\usepackage{graphicx}
\usepackage{graphics}
\usepackage{epsfig}
\def\beq{\begin{equation}}
\def\eeq{\end{equation}}
\def\bea{\begin{eqnarray}}
\def\eea{\end{eqnarray}}

\def\bec{\begin{center}}
\def\eec{\end{center}}
\def\nn{\nonumber}

\def\cQ{{\cal Q}}

\begin{document}
\small
\preprint{SU-4252-871 \vspace{1cm}} \setlength{\unitlength}{3mm}
\title{Lattice actions for Yang-Mills quantum mechanics with exact
supersymmetry\vspace{0.5cm}}
\author{Simon Catterall$^{a}$}\thanks{smc@phy.syr.edu}\author{ Anosh Joseph$^{a}$}\thanks{ajoseph@phy.syr.edu}

\affiliation{$^{a}$Department of Physics, Syracuse University,
Syracuse, NY 13244-1130, USA}
\vspace{1.0cm}
\begin{abstract}
\vspace{0.5cm}
We derive lattice actions for Yang-Mills quantum mechanics for models with
$\cQ=4, 8$ and $16$ supercharges which possess an exact supersymmetry at
non-zero lattice spacing. These are obtained by dimensional reduction 
of twisted versions of the corresponding
super Yang-Mills theories in $D=2, 3$ and $4$ dimensions. 
\end{abstract}
\maketitle
\section{INTRODUCTION}

Supersymmetric Yang-Mills theories are interesting both as playgrounds for
understanding quantum field theory and as gauge theories which are conjectured
to be dual to certain string theories \cite{Maldacena, Itzhaki}.
Typically these dualities between string and gauge theory require that
the gauge theory be taken at strong coupling. This requirement motivates
defining the theory on a lattice which would allow for strong coupling
expansions and Monte Carlo simulation. Perhaps the simplest of these
gauge-gravity dualties is exhibited by the conjectured equivalence of
the type IIA string theory containing $N$ $D0$-branes and super
Yang-Mills quantum mechanics with gauge group $SU(N)$. More specifically
the large $N$ limit of the gauge model at low
temperature $T$ or strong `t Hooft coupling is thought to
provide a description of the black hole that arises in the low
energy supergravity
limit of the string theory. Initial investigations of this and related
models have been reported in the literature \cite{Nishimura, Wiseman, Jun}.
The numerical work so far has employed actions that do not possess
an exact supersymmetry \cite{Comment1}. The purpose of this work is to 
derive lattice actions
which retain an exact supersymmetry at non-zero
lattice spacing which could be used in similar
Monte Carlo studies.

Unfortunately, conventional discretizations of supersymmetric theories
break supersymmetry completely and the resultant lattice theories typically
require a great deal of fine tuning in order that supersymmetry is recovered
in the continuum limit \cite{Simon1, Kaplan, Feo}. However, it has been
shown that in theories with a multiple of $2^D$ supercharges, where $D$ is the total number of spacetime dimensions, 
one or more supersymmetries can be retained provided the discretization
scheme is chosen carefully. Two approaches have been successfully pursued
based on either orbifolding a supersymmetric matrix model
\cite{orbifold,Damgaard0,Giedt_rev} or direct
discretization of a reformulation of the continuum supersymmetric theory
in terms of so-called {\it twisted} variables
\cite{top,Sugino1,Simon2,Dadda,Dadda2}. Recently, these two
approaches have been shown to be equivalent \cite{Unsal,Damgaard1, Damgaard2,
Takimi, new}.

The idea of twisting goes back to the seminal paper of
Witten in which the twisted formulation was used to
construct a topological field theory \cite{Witten}. The process
naturally exposes a nilpotent scalar supersymmetry with the topological
sector of the twisted theory corresponding to operators invariant under the
action of this scalar supersymmetry. In the context of 
creating supersymmetric
lattice theories this projection to the $\cQ$-invariant subspace is
dropped and the twisting (in flat space) is simply regarded as a 
convenient change of
variables - one more suitable for discretization.
Indeed, the fermionic content is then encoded by a series of antisymmetric
tensor fields which can be embedded as components of
one or more K\"{a}hler-Dirac fields
\cite{Kawamoto}. As was shown by Rabin \cite{Rabin},
theories
involving K\"{a}hler-Dirac fields may be discretized without inducing
fermion doubling problems and indeed at the level of free field theory
the resultant lattice theories are equivalent to staggered fermions \cite{Banks}.

In addition to a geometrical treatment of the fermions the twisted
formulation has the merit of allowing the action to be written in
$\cQ$-exact form. Thus the problem of 
translating the $\cQ$-invariance of the continuum
theory to the lattice is replaced by the simpler requirement of
keeping the scalar supercharge nilpotent when acting on the
lattice fields. Typically this is a much simpler proposition and is the
one adopted by all discretizations of the twisted theories 
considered so far \cite{Sugino1,
Sugino2,Simon2,Simon3,Dadda,Dadda2,new}. 

In this paper, we start from the twisted forms of the gauge theories 
in two, three and four dimensions which possess $\cQ=4, 8$ and $16$ supersymmetries
and dimensionally reduce them to one (Euclidean) dimension. The
resultant continuum theories can be written in terms of multiples
of the basic K\"{a}hler-Dirac field which in one dimension
which contains one scalar and one vector field. We show also that each new
scalar fermion is associated with an additional 
scalar supersymmetry which is inherited from the dimensional reduction. 
Discretization
then proceeds using a prescription due to Sugino \cite{Sugino2}. 
\section{The four supercharge model}
Consider the continuum twisted form of the two dimensional ${\cal N}=2$ (Euclidean) super Yang-Mills model given in eg.
\cite{Simon2}. The bosonic part of this theory 
contains two scalar fields $\phi$, $\bar{\phi}$, a vector $A_{\mu}$ and a 
tensor field $B_{\mu \nu}$. The fermionic part consists of an anticommuting scalar field $\eta$, a vector $\psi_{\mu}$ and a field $\chi_{\mu \nu}$. If $\cQ$ is a scalar supercharge obtained by {\it twisting} the original Majorana supercharges of the theory, the $\cQ$-variation of the gauge fermion
\beq
\label{eq:Lambda2}
\Lambda = \textrm{Tr} \int d^{2}x \; \Big(\frac{1}{4}\eta[\phi, \bar{\phi}] +\chi_{\mu \nu}F_{\mu \nu} + \frac{1}{2}\chi_{\mu \nu}B_{\mu \nu} + \psi_{\mu}D_{\mu}\bar{\phi}\Big)
\eeq 
will give us a twisted action
\beq
S = \beta \cQ \Lambda,
\eeq
where $\beta$ is a coupling constant.
Dimensional reduction of the action to one dimension will then yield a supersymmetric Yang-Mills quantum mechanics theory. Dimensionally reducing eqn. (\ref{eq:Lambda2}) with respect to the $x^{2}$-direction we find
\beq
\label{eq:Lambda2oneD}
\Lambda = \textrm{Tr} \int dx \; \Big(\frac{1}{4}\eta[\phi, \bar{\phi}] +2\chi_{12}D_{1}A_{2} + \chi_{12}B_{12} + \psi_{1}D_{1}\bar{\phi}+ \psi_{2}[A_{2},\bar{\phi}]\Big).
\eeq 

The scalar supercharge $\cQ$ acts on the component fields as follows
\bea
\cQ \; A_{1} &=& \psi_{1}, \nn \\
\cQ \; A_{2} &=& \psi_{2}, \nn \\
\cQ \; \psi_{1} &=& - D_{1}\phi, \nn \\
\cQ \; \psi_{2} &=& -[A_{2}, \phi], \nn \\
\cQ \; \phi &=& 0, \\
\cQ \; \chi_{12} &=& B_{12},\nn \\
\cQ \; B_{12} &=& [\phi, \chi_{12}], \nn \\
\cQ \; \bar{\phi} &=& \eta, \nn \\
\cQ \; \eta &=& [\phi, \bar{\phi}]. \nn
\eea
Carrying out the $\cQ$-variation on eqn. (\ref{eq:Lambda2oneD}) and integrating over the multiplier field $B_{12}$ we arrive at the action
\bea
\label{eq:action1d}
S &=& \beta \; \textrm{Tr} \int dx \; \Big(-\sum_{i=1}^{3}(D_{1}\phi^{i})^{2} - \sum_{i<j, \; i,j = 1}^{3} \; [\phi^{i}, \phi^{j}]^{2} -2\psi_{1}^{1}D_{1}\eta^{1}- 2\psi_{1}^{2}D_{1}\eta^{2} +\psi_{1}^{1}[\bar{\phi}, \psi_{1}^{1}]\nn \\
&&~~~~~~~~~~~~~~~~~-\psi_{1}^{2}[\phi, \psi_{1}^{2}]-\eta^{1}[\phi, \eta^{1}]+\eta^{2}[\bar{\phi}, \eta^{2}] + 2\psi_{1}^{1}[\phi^{3}, \psi_{1}^{2}] -2 \eta^{1}[\phi^{3},\eta^{2}]\Big).
\eea

Notice that we have relabelled the fields:
\bea
\textrm{\small{OLD}} &\rightarrow& \textrm{\small{NEW}}\nn \\
\psi_{1} &\rightarrow& \psi_{1}^{1}\nn \\
\chi_{12} &\rightarrow& \psi_{1}^{2}\nn \\
\eta /2 &\rightarrow& \eta^{1}\nn \\
\psi_{2} &\rightarrow& \eta^{2}\nn \\
A_{2} &\rightarrow& \phi^{3}\nn
\eea 
It will also prove convenient to decompose the scalar fields $\phi = -\phi^{1}+i \phi^{2}$ and $\bar{\phi} =-\phi^{1} -i \phi^{2}$. 

In terms of the relabelled fields we write the part of the action comprising the fermionic and Yukawa terms in the form
\beq
S_{F} + S_{Y} = \int dx \; \textrm{Tr} \; \Psi^{\dagger}\Gamma^{4} D_{1}\Psi + \sum_{i=1}^{3}\; \Psi^{\dagger}[\Gamma^{i} \phi^{i}, \Psi],
\eeq
where 
\beq
\Gamma^{1}={\bf 1} \otimes \sigma_{3},~~\Gamma^{2}=-i~{\bf 1} \otimes {\bf 1},~~\Gamma^{3}=\sigma_{3} \otimes \sigma_{1},~~\Gamma^{4}=\sigma_{1} \otimes {\bf 1},
\eeq
and the spinor
\beq
\Psi^{\dagger} =(\psi_{1}^{1} \; \psi_{1}^{2} \; \eta^{1} \; \eta^{2})~.
\eeq

This form of the twisted theory can be related to the usual
action for ${\cal N}=1$ super Yang-Mills theory reduced to one dimension
by recognizing that the usual 4d Majorana matrices given by
\beq
\gamma^{1}=i \sigma_{1} \otimes \sigma_{3},~~\gamma^{2}= i~{\bf 1} \otimes {\bf 1},~~\gamma^{3}=-i \sigma_{1} \otimes \sigma_{1},~~\gamma^{4}=-i \gamma^{0}=i \sigma_{1} \otimes \sigma_{2}
\eeq
may be transformed to the above $\Gamma$ representation using the similarity 
transformation \footnote{Thanks to Toby Wiseman for
pointing out the correct similarity transformation}
\beq
\Gamma^{1} = i S \gamma^{4} \gamma^{3} S^{-1},~~\Gamma^{2} = i S \gamma^{4} \gamma^{4} S^{-1},~~\Gamma^{3} = i S \gamma^{4} \gamma^{1} S^{-1},~~\Gamma^{4} = i S \gamma^{4} \gamma^{2} S^{-1}, 
\eeq
where
\bea
&&S = \left( \begin{array}{cc}
\sigma_{1} &  0 \\
0 & \sigma_{3} \end{array} \right).
\eea

We see that the twisted fermions fill out the usual 4d Majorana spinor
as expected.

Notice that the twisted theory is actually
equivalent to a dimensional reduction of the
usual ${\cal N}=1$ theory in four dimensional Minkowski space 
along one space and the {\it time} direction followed by a Wick
rotation of the original temporal direction. The latter
corresponds to $A^{\rm Min}_t\to \phi^{\rm Min}_2\to i\phi^{\rm Eucl}_2$.
The final Euclidean time direction is then associated with the
Dirac matrix
$\Gamma^{4}=\left(\begin{array}{cc}0&I\\I&0\end{array}\right)$. In this
form it is easily discretized as we will see later while maintaining the
antisymmetry of the discrete Dirac operator.

The twist decomposition of ${\cal N}=2$ theory gives rise
to four supercharges - two scalars and two vectors. 
We have used only one scalar supercharge $\cQ$ in deriving the above 
continuum action from the gauge fermion $\Lambda$. 
Clearly the new scalar fermion that appears in the dimensionally reduced theory
is related to the existence of this second scalar supersymmetry.
Its action on the fields can be uncovered by
noticing that 
the continuum action given in eqn. (\ref{eq:action1d}) 
is invariant under the field transformations
\bea
\psi_{1}^{1} &\rightarrow& \psi_{1}^{2}\nn \\
\eta^{1} &\rightarrow& \eta^{2}\\ 
\phi^3 &\rightarrow& \phi^3\nn\\
\phi~ &\rightarrow& -\overline{\phi}\nn
\eea
This set of field transformations is a symmetry of the continuum action. 
We can combine this with the 
action of the scalar supercharge $\cQ$ to derive an additional 
supersymmetry, say $\cQ'$ of the theory. 
%% edit, simon mar29 2008
It corresponds to the vector supercharge $Q_2$ of the two dimensional
parent theory before dimensional reduction. In the continuum the theory
can then also be written in a $\cQ'$-exact form. However, we will see that
in our lattice construction this second supersymmetry is broken by terms
of order the lattice spacing. However, using 
the arguments given in \cite{Wiseman} it should
be recovered without fine tuning in the continuum limit.  
This supersymmetry will transform component fields of the continuum 
theory in the following way
\bea
&&\cQ'~A_{1} = \psi_{1}^{2}, \nn \\
&&\cQ'~\phi^3 = \eta^{1}, \nn \\
&&\cQ'~\psi_{1}^{1} = B_{12}, \nn \\
&&\cQ'~\psi_{1}^{2} = D_1\bar{\phi}, \nn \\
&&\cQ'~\eta^{1} = -[\bar{\phi},\phi^3], \\
&&\cQ'~\eta^{2} = -\frac{1}{2}[\phi,\bar{\phi}],\nn \\
&&\cQ'~\bar{\phi} = 0, \nn\\
&&\cQ'~\phi = -2 \eta^{2}, \nn\\
&&\cQ'~B_{12}= -[\bar{\phi}, \psi_{1}^{1}]\nn
\eea
This method for deriving additional twisted supersymmetries 
is described in some
detail in \cite{rest}.

Turning now to the lattice, it is straightforward to discretize this theory
in a way which preserves the nilpotency of $\cQ$.
The prescription was given by Sugino \cite{Sugino1}
in the context of the two dimensional model. Here, we trivially
extend it to discretization of a one dimensional model. We place
all the fields on sites of a regular one dimensional
lattice except the gauge field $A_{1}(x)$, which is replaced by 
a unitary variable $U_{1}(x)$ living on the link $(x, x+\hat{1})$. 
We write the lattice gauge fermion $\Lambda$ in terms of lattice variables
\bea
\label{eq:LambdaLattice}
\Lambda_{\textrm{\tiny{L}}} &=& \textrm{Tr} \sum_{x} \Big(~\frac{1}{4}\eta(x)[\phi(x), \bar{\phi}(x)] +2\chi_{12}(x)D_{1}^{+}A_{2}(x) + \chi_{12}(x)B_{12}(x) + \psi_{1}(x) D_{1}^{+}\bar{\phi} \nn \\
&&~~~~~~~~+\psi_{2}(x)[A_{2}(x),\bar{\phi}(x)]~\Big),
\eea
where the forward difference operator is defined as
\beq
D_{\mu}^{+} f(x) = U_{\mu}(x) f(x + \mu)U^\dagger_{\mu}(x) - f(x).
\eeq 
The scalar supersymmetry transformation rules on the lattice take the form
\cite{Sugino1}
\bea
\cQ \; U_{1}(x) &=& \psi_{1}(x)U_{1}(x), \nn \\
\cQ \; A_{2}(x) &=& \psi_{2}(x), \nn \\
\cQ \; \psi_{1}(x) &=& \psi_{1}(x)\psi_{1}(x) - D_{1}^{+}\phi(x), \nn \\
\cQ \; \psi_{2}(x) &=& -[A_{2}(x), \phi(x)], \nn \\
\cQ \; \phi(x) &=& 0, \\
\cQ \; \chi_{12}(x) &=& B_{12}(x), \nn \\
\cQ \; B_{12}(x) &=& [\phi(x), \chi_{12}(x)], \nn \\
\cQ \; \bar{\phi}(x) &=& \eta(x), \nn \\
\cQ \; \eta(x) &=& [\phi(x), \bar{\phi}(x)]. \nn
\eea
These transformations reduce to their continuum counterparts in the limit
of vanishing lattice spacing where, for example, the term quadratic
in $\psi_\mu$ is suppressed by an additional power of the lattice spacing.
Notice that $\cQ^2$ is still nilpotent up to lattice gauge
transformations. In dimensions two or greater this prescription is
problematic in that it leads to a gauge action which possesses many
vacua all but one of which are absent in the continuum
theory. However, in one dimension this term is missing and the
lattice prescription is well defined.

Applying the lattice $\cQ$-variation to eqn. (\ref{eq:LambdaLattice}) and integrating out the multiplier fields we get the lattice action
\beq
S = S_{B} + S_{F} + S_{Y} + S_{R}~,
\eeq
where the bosonic part of the action 
\beq
S_{B} = \beta \; \textrm{Tr}\; \sum_{x}\Big(-\sum_{i=1}^{3}(D_{1}^{+}\phi^{i}(x))^{2} - \sum_{i<j, \; i,j = 1}^{3} \; [\phi^{i}(x), \phi^{j}(x)]^{2}\Big)~,
\eeq
the fermionic kinetic term is
\beq
S_{F} = \beta \; \textrm{Tr} \sum_{x}\; -2\Big(\psi_{1}^{1}D_{1}^{+}\eta^{1}(x) + \psi_{1}^{2}(x)D_{1}^{+}\eta^{2}(x)\Big)~,
\eeq
and the Yukawa part
\bea
S_{Y} &=& \beta \; \textrm{Tr}\; \sum_{x}~\Big(~\psi_{1}^{1}(x)[\widetilde{\bar{\phi}}(x), \psi_{1}^{1}(x)] + 2\psi_{1}^{1}(x)[\widetilde{\phi}^{3}(x), \psi_{1}^{2}(x)] - \psi_{1}^{2}(x)[\phi(x), \psi_{1}^{2}(x)]  \nn \\
&&~~~~~~~~~~~-\eta^{1}(x)[\phi(x), \eta^{1}(x)] - 2 \eta^{1}(x)[\phi^{3}(x),\eta^{2}(x)] + \eta^{2}(x)[\bar{\phi}(x), \eta^{2}(x)]\Big)~.
\eea
In addition, the lattice action picks up a residual part
\beq
S_{R} = -\beta \; \textrm{Tr}\; \sum_{x} \psi_{1}^{1}(x)\psi_{1}^{1}(x)D_{1}^{+}\phi^{1}(x) + i\psi_{1}^{1}(x)\psi_{1}^{1}(x)D_{1}^{+}\phi^{2}(x)~.
\eeq
Here also we have rescaled and relabelled the fields as in the continuum case. Notice that the fields $\phi$ and $\bar{\phi}$ appear in an unusual way in the Yukawa part of the action: 
\beq
\widetilde{\phi}(x) = U_{1}(x)\phi(x +\hat{1})U_{1}(x)^{\dagger}\nn
\eeq
and
\beq
\widetilde{\bar{\phi}}(x) = U_{1}(x)\bar{\phi}(x+\hat{1})U_{1}^{\dagger}(x).\nn
\eeq
This smearing of certain scalar Yukawa terms 
is due to the non-trivial $\cQ$-transformations 
we defined on $U_{1}(x)$ and $\psi_{1}(x)$. The 
residual part of the lattice action is also a 
consequence of these non-trivial transformations. Furthermore, 
the presence of these
point smeared Yukawas and this residual piece in the lattice action
break the $\cQ'$-symmetry introduced earlier.

The fermion kinetic term can be expressed in the form 
\beq
{\bf D}_{1} = \left( \begin{array}{cc}
0 & ({\mathbb I}_{2 \times 2})D_{1}^{+}\\
({\mathbb I}_{2 \times 2})D_{1}^{-} & 0 \end{array} \right)
\eeq
which has the same $\Gamma^{4}$ structure as
appeared in the continuum case and is explicitly
antisymmetric as required. However, the Yukawa interactions cannot be
put in such a simple form 
as a result of the appearance of
terms depending on the smeared scalar fields
as described above. Instead the Yukawa part takes the form
\beq
{\bf \Psi^{\dagger}[\tilde{O}, \Psi]},
\eeq
where the matrix
\beq
{\bf \tilde{O}} = \left( \begin{array}{cccc}
\widetilde{\bar{\phi}} & \widetilde{\phi}^{3} & 0 & 0 \\
\widetilde{\phi}^{3} & -\phi & 0 & 0  \\
0 & 0 & \bar{\phi} & -\phi^{3}  \\
0 & 0 & -\phi^{3} & -\phi \end{array} \right)
\eeq
\section{The Eight Supercharge Model}
Here we consider the twisted version of ${\cal N} = 4$ super Yang-Mills theory in three (Euclidean) dimensions. The theory contains eight supercharges. The bosonic
part consists of a gauge potential $A_{\mu}$, where $\mu=1, 2, 3$, two scalars $\phi$ and $\bar{\phi}$, and two
fields $B_{\mu \nu}$ and $W_{\mu \nu \lambda}$. The fermionic part of the theory consists of one anti-commuting scalar $\eta$, a vector $\psi_{\mu}$, a tensor $\chi_{\mu\nu}$
and the 3-form field $\theta_{\mu \nu \lambda}$.

The twisted action can again be written in a $Q$-exact form
\beq
S = \beta~Q \Lambda,
\eeq
where the gauge fermion $\Lambda$ takes the form
\bea
\Lambda &=& \int d^{3}x \; \textrm{Tr}\; \Big( \chi_{\mu \nu}(F_{\mu \nu} + \frac{1}{2}B_{\mu \nu} + D_{\lambda}W_{\lambda \mu \nu}) + \psi_{\mu}D_{\mu}\bar{\phi}+ \frac{1}{4}\eta[\phi, \bar{\phi}] + \frac{1}{3!} \theta_{\mu \nu \lambda}[W_{\mu \nu \lambda}, \bar{\phi}]\Big).
\eea

Dimensional reduction of this action to one dimension will give a supersymmetric quantum mechanics with eight supercharges. The gauge fermion $\Lambda$, after dimensionally reducing along $x^{2}$- and $x^{3}$-directions
\bea
\Lambda &=& \int d^{3}x \; \textrm{Tr}\; \Big( \chi_{12}D_{1}A_{2} + \chi_{13}D_{1}A_{3} + \chi_{23}[A_{2}, A_{3}] + \chi_{12}B_{12} +\chi_{13}B_{13}+\chi_{23}B_{23}\nn \\
&&~~~~~~~~~~+ \chi_{12}[A_{3}, W_{312}] + \chi_{13}[A_{2}, W_{213}] + \chi_{23}D_{1}W_{123} + \psi_{1}D_{1}\bar{\phi}+ \psi_{2}[A_{2}, \bar{\phi}] \nn \\
&&~~~~~~~~~~+ \psi_{3} [A_{3}, \bar{\phi}]+ \frac{1}{4}\eta[\phi, \bar{\phi}] + \theta_{123}[W_{123}, \bar{\phi}]\Big).~~
\eea
Again it is straightforward to we write down the scalar supercharge transformation rules for the component fields
\bea
&& Q A_{\mu} = \psi_{\mu},\nn \\
&& Q \psi_{1} = - D_{1}\phi, \nn \\
&& Q \psi_{i} = - [A_{i}, \phi],~ i \neq 1, \nn \\
&& Q \phi = 0, \nn \\
&& Q \bar{\phi} = \eta,\nn \\
&& Q \eta = [\phi, \bar{\phi}], \\
&& Q B_{\mu \nu} = [\phi, \chi_{\mu \nu}],\nn \\
&& Q \chi_{\mu \nu} = B_{\mu \nu},\nn \\
&& Q W_{\mu \nu \lambda} = \theta_{\mu \nu \lambda},\nn \\
&& Q \theta_{\mu \nu \lambda} = [\phi, W_{\mu \nu \lambda}]. \nn
\eea 

After integrating out the multiplier field $B_{\mu \nu}$, and using the Bianchi identity, the bosonic part of the action can be written in the following form
\bea
S_{B} &=& \beta \int dx\; \textrm{Tr} ~ \Big(-(D_{1}A_{2})^{2}-(D_{1}A_{3})^{2}-[A_{2}, A_{3}]^{2} - (D_{1}W_{123})^{2}-[A_{2},W_{231}]^{2}- [A_{3}, W_{312}]^{2}\nn \\
&&~~~~~~~~~~~~~~- (D_{1}\phi^{1})^{2}- (D_{1}\phi^{2})^{2} -[A_{2}, \phi^{1}]^{2} - [A_{2}, \phi^{2}]^{2}-[A_{3},\phi^{1}]^{2} - [A_{3}, \phi^{2}]^{2}\nn \\
&&~~~~~~~~~~~~~~-[\phi^{1},\phi^{2}]^{2} - [\phi^{1}, W_{123}]^{2} - [\phi^{2}, W_{123}]^{2}\Big)
\eea
where we have decomposed the fields $\phi = \phi^{1} +i \phi^{2}$ and $\bar{\phi} = \phi^{1} -i \phi^{2}$.

Relabelling the fields
\bea
\phi^{3} &=& A_{2}\nn \\
\phi^{4} &=& A_{3}\\
V^{1} &=& W_{123}\nn
\eea
the bosonic part of the action becomes
\beq
S_{B} = \beta \int dx\; \textrm{Tr} ~~-\sum_{i=1}^{4} (D_{1}\phi^{i})^{2} -(D_{1}V^{1})^{2}- \sum_{i<j;\; i, j =1}^{4} [\phi^{i}, \phi^{j}]^{2}- \sum_{i =1}^{4}~[\phi^{i}, V^{1}]^{2}.
\eeq

Dimensional reduction of the fermionic kinetic part of the action will give pure kinetic part $S_{F}$ corresponding to the
$x^{1}$-direction together with Yukawa couplings $S_{FY}$ from the $x^{2}$- and $x^{3}$-directions
\bea
S_{F} &=& -2 \beta \int dx\; \textrm{Tr} \Big[\chi_{12}D_{1}\psi_{2}+\chi_{13}D_{1}\psi_{3}+\chi_{23}D_{1}\theta_{123}+\frac{\eta}{2} D_{1} \psi_{1}\Big],
\eea
\bea
S_{FY} &=& 2 \beta \int dx\; \textrm{Tr}\; \Big[\chi_{12}[A_{2},\psi_{1}]-\chi_{23}[A_{2},\psi_{3}]+\chi_{13}[A_{2},\theta_{123}]-\frac{\eta}{2}[A_{2}, \psi_{2}]+\chi_{13}[A_{3},\psi_{1}]\nn \\
&&~~~~~~~~~~~~~~+\chi_{23}[A_{3},\psi_{2}]-\chi_{12}[A_{3},\theta_{123}]-\frac{\eta}{2} [A_{3},\psi_{3}]\Big].~~
\eea
The Yukawa part of the action being
\bea
S_{Y} &=& \beta \int dx \; \textrm{Tr} \Big(-\frac{\eta}{2}[\phi, \frac{\eta}{2}]-\chi_{12}[\phi, \chi_{12}]-\chi_{13}[\phi, \chi_{13}]-\chi_{23}[\phi, \chi_{23}] + \psi_{1}[\bar{\phi}, \psi_{1}]\nn \\
&&~~~~~~~~~~~~~+ \psi_{2}[\bar{\phi}, \psi_{2}]+ \psi_{3}[\bar{\phi}, \psi_{3}]+ \theta_{123}[\bar{\phi}, \theta_{123}]+2 \frac{\eta}{2}[\theta_{123}, W_{123}]+2 \chi_{12}[W_{123}, \psi_{3}]\nn \\
&&~~~~~~~~~~~~~-2 \chi_{13}[W_{123}, \psi_{2}]+2 \chi_{23}[W_{123}, \psi_{1}]\Big).
\eea

Relabelling the fermionic fields
\bea
OLD &\longrightarrow& NEW \nn \\
-\psi_{1} &\longrightarrow& \psi_{1}^{1}\nn \\
\chi_{12} &\longrightarrow& \psi_{1}^{2}\nn \\
\chi_{13} &\longrightarrow& \psi_{1}^{3}\nn \\
\theta_{123} &\longrightarrow& \psi_{1}^{4}\nn \\
-\frac{\eta}{2} &\longrightarrow& \eta^{1}\nn \\
\psi_{2} &\longrightarrow& \eta^{2}\nn \\
\psi_{3} &\longrightarrow& \eta^{3}\nn \\
\chi_{23} &\longrightarrow& \eta^{4}\nn 
\eea
the action for the supersymmetric Yang-Mills quantum mechanics can be written in a more compact form
\beq
S = \beta \; \int dx \; \textrm{Tr}\; \Psi^{\dagger}\; \Gamma^{6}D_{1} \Psi + \sum_{i=1}^{4}\Psi^{\dagger}[\Gamma^{i} \phi^{i}, \Psi] + \Psi^{\dagger}[\Gamma^{5} V^{1}, \Psi],
\eeq
where the spinor 
\beq
\label{eq:Spinor8}
\Psi^{\dagger} =(\psi_{1}^{1} \; \psi_{1}^{2} \; \psi_{1}^{3} \; \psi_{1}^{4} \; \eta^{1} \; \eta^{2} \; \eta^{3} \; \eta^{4}),
\eeq
and the $\Gamma$'s
\bea
&&\Gamma^{1} = \sigma_{3} \otimes \sigma_{3} \otimes \sigma_{3},~~~\Gamma^{2} = -i~1 \otimes 1 \otimes 1,~~~~\Gamma^{3} = -\sigma_{3} \otimes \sigma_{3} \otimes \sigma_{1},\nn \\
&&\Gamma^{4} = -\sigma_{3} \otimes \sigma_{1} \otimes 1,~~\Gamma^{5} = \sigma_{2} \otimes \sigma_{1} \otimes \sigma_{2},~~~~\Gamma^{6} = -\sigma_{1} \otimes 1 \otimes 1.\nn
\eea

Dimensional reduction of the eight supercharge action leads to four scalar fermions and thus four scalar supercharges. So far we have made use of only one of them, the supercharge $\cQ$, to construct the action. The remaining scalar supersymmetries may be revealed using the scalar supercharge $\cQ$ and additional symmetries of the action. For example, to obtain another scalar supersymmetry say $\cQ'$, we look at the invariance of the action under the field transformations

\bea
&& V^{1} \rightarrow -V^{1}, ~~~~~~~~\phi^{3} \rightarrow -\phi^{3},
\eea
along with
\bea
&&\psi_{1}^{1} \rightarrow \psi_{1}^{4},~~~~~~~~~~~~~~~~ \eta^{1} \rightarrow \eta^{4},\nn \\
&&\psi_{1}^{2} \rightarrow \psi_{1}^{3},~~~~~~~~~~~~~~~~ \eta^{2} \rightarrow \eta^{3},\nn \\
&&\psi_{1}^{3} \rightarrow \psi_{1}^{2},~~~~~~~~~~~~~~~~ \eta^{3} \rightarrow \eta^{2}, \\
&&\psi_{1}^{4} \rightarrow \psi_{1}^{1},~~~~~~~~~~~~~~~~ \eta^{4} \rightarrow \eta^{1}.\nn
\eea

The scalar supersymmetry, $\cQ'$ associated with this invariance of the action is
\bea
&&\cQ' \psi_{1}^{1}= [\phi, V^{1}],~~~~~~~~\cQ' \eta^{1}=B_{23},~~~~~~~~~\cQ' B_{12}=[\phi, \psi_{1}^{3}],~~~~~~~~~~\cQ' \phi=0,\nn \\
&&\cQ' \psi_{1}^{2}= B_{13},~~~~~~~~~~~~\cQ' \eta^{2}=[\phi, \phi^{4}],~~~~~~\cQ' B_{13}=[\phi, \psi_{1}^{2}],~~~~~~~~~~\cQ' \bar{\phi}=-\eta^{4},\nn \\
&&\cQ' \psi_{1}^{3}= B_{12},~~~~~~~~~~~~\cQ' \eta^{3}=-[\phi, \phi^{3}],~~~~\cQ' B_{23}=[\phi, \eta^{1}],~~~~~~~~~~\cQ' \phi^{3}= -\eta^{3},\nn \\
&&\cQ' \psi_{1}^{4}= D_{1}\phi,~~~~~~~~~~~\cQ' \eta^{4}=[\phi, \bar{\phi}],~~~~~~~~\cQ' A_{1}=\psi_{1}^{4},~~~~~~~~~~~~~~~~\cQ' \phi^{4}= \eta^{2},\nn \\
&&\cQ' V^{1}=\psi_{1}^{1}.\nn
\eea

The lattice action is obtained by the $\cQ$-variation of the gauge fermion
\bea
\Lambda &=& \sum_{x} \; \textrm{Tr}\; \Big( \chi_{12}(x)D^{+}_{1}A_{2}(x) + \chi_{13}(x)D^{+}_{1}A_{3}(x) + \chi_{23}(x)[A_{2}(x), A_{3}(x)] + \chi_{12}(x)B_{12}(x) \nn \\
&&~~~~~~~~~~+\chi_{13}(x)B_{13}(x)+\chi_{23}(x)B_{23}(x)+ \chi_{12}(x)[A_{3}(x), W_{312}(x)] + \chi_{13}(x)[A_{2}(x), W_{213}(x)] \nn \\
&&~~~~~~~~~~+ \chi_{23}(x)D^{-}_{1}W_{123}(x)  + \psi_{1}(x)D^{+}_{1}\bar{\phi}(x)+ \psi_{2}(x)[A_{2}(x), \bar{\phi}(x)]+ \psi_{3}(x)[A_{3}(x), \bar{\phi}(x)]\nn \\
&&~~~~~~~~~~+ \frac{1}{4}\eta(x)[\phi(x), \bar{\phi}(x)] + \theta_{123}(x)[W_{123}(x), \bar{\phi}(x)]\Big),~~
\eea
where the backward difference operator is defined as
\beq
D_{\mu}^{-}g_{\mu}(x)=g_{\mu}(x)-U_{\mu}^{\dagger}(x-\mu)g_{\mu}(x-\mu)U_{\mu}(x-\mu),
\eeq
and the $\cQ$-transformation rules for fields on lattice are similar to the $\cQ=4$ model
\bea
&&\cQ U_{1}(x) = \psi_{1}(x)U_{1}(x)~~~~~~~~~~~~~~~~~~~~~~~~\cQ A_{\mu}(x) = \psi_{\mu}(x)\nn \\
&&\cQ \psi_{1}(x) = \psi_{1}(x)\psi_{1}(x) - D_{1}^{+}\phi(x),~~~~~~~~~\cQ \psi_{i}(x) = -[A_{i}(x), \phi(x)], \; \; i \neq 1,\nn \\
&&\cQ \eta(x) = [\phi(x), \bar{\phi}(x)],~~~~~~~~~~~~~~~~~~~~~~~~\cQ B_{\mu \nu}(x) = [\phi(x), \chi_{\mu \nu}(x)]\nn \\
&&\cQ \chi_{\mu \nu}(x) = B_{\mu \nu}(x),~~~~~~~~~~~~~~~~~~~~~~~~~~~\cQ W_{\mu \nu \lambda}(x) = \theta_{\mu \nu \lambda}(x) \\
&&\cQ \theta_{\mu \nu \lambda}(x) = [\phi(x), W_{\mu \nu \lambda}(x)],~~~~~~~~~~~~~~~~\cQ \phi(x) = 0,\nn \\
&&\cQ \bar{\phi}(x) = \eta(x).\nn
\eea 

Carrying out the $\cQ$-variation we write down the action

\beq
S = S_{B} + S_{F} + S_{Y}+S_{R}.
\eeq

In terms of the relabelled fields

\bea
&&\psi_{1}^{1}(x) = -\psi_{1}(x),~~~~~~~~~~~~~\eta^{1}(x) = -\frac{\eta}{2}(x), \nn \\
&&\psi_{1}^{2}(x) = \chi_{12}(x),~~~~~~~~~~~~~~\eta^{2}(x) = \psi_{2}(x), \nn \\
&&\psi_{1}^{3}(x) = \chi_{13}(x),~~~~~~~~~~~~~~\eta^{3}(x) = \psi_{3}(x), \nn \\
&&\psi_{1}^{4}(x) = \theta_{123}(x),~~~~~~~~~~~~~~\eta^{4}(x) = \chi_{23}(x),\nn \\
&&\phi^{3}(x) = A_{2}(x),~~~~~~~~~~~~~~~~\phi^{4}(x) = A_{3}(x), \nn \\
&&V^{1}(x) = W_{123}(x),~~~~~~~~~~~~~\phi(x) = \phi^{1}(x) + i \phi^{2}(x).\nn
\eea

the bosonic part takes the form
\bea
S_{B} &=& \beta \sum_{x}\; \textrm{Tr} \Big[-\Big(D^{+}_{1}\phi^{3}(x)+D^{+}_{1}\phi^{4}(x)+[\phi^{3}(x), \phi^{4}(x)]\Big)^{2} - \Big(D^{-}_{1}V^{1}(x) - [\phi^{3}(x), V^{1}(x)] \nn \\
&&~~~~~~~~~~~~~~~+ [\phi^{4}(x), V^{1}(x)]\Big)^{2}-\Big(D^{+}_{1}\phi^{1}(x)\Big)^{2} - \Big(D^{+}_{1}\phi^{2}(x)\Big)^{2}- [\phi^{3}(x), \phi^{1}(x)]^{2}\nn \\
&&~~~~~~~~~~~~~~- [\phi^{3}(x), \phi^{2}(x)]^{2}- [\phi^{4}(x), \phi^{1}(x)]^{2} - [\phi^{4}(x), \phi^{2}(x)]^{2} - [\phi^{1}(x), \phi^{2}(x)]^{2}\nn \\
&&~~~~~~~~~~~~~~-[\phi^{1}(x), V^{1}(x)]^{2} - [\phi^{2}(x), V^{1}(x)]^{2}\Big].
\eea

Notice that unlike the continuum case this cannot be further simplified
since the lattice theory does not possess an exact Bianchi identity.
The fermionic kinetic term can be expressed in a matrix form similar to the $\Gamma^{6}$ matrix in the continuum, with the same spinor structure as eqn. (\ref{eq:Spinor8}),
\beq
{\bf D}_{1} = \left( \begin{array}{cc}
0 &  ({\mathbb I}_{4 \times 4})D_{1}^{-} \\
({\mathbb I}_{4 \times 4})D_{1}^{+} & 0 \end{array} \right)
\eeq

The Yukawa part of the action $S_{Y}$ can also be expressed in a matrix form
\beq
{\bf \Psi^{\dagger}[\tilde{O}, \Psi]},
\eeq
where
\beq
{\bf \tilde{O}} = \left( \begin{array}{cccccccc}
\tilde{\bar{\phi}} & -\tilde{\phi}^{3} & -\tilde{\phi}^{4} & 0 & 0 & 0 & 0 & -\tilde{V}^{1} \\
-\tilde{\phi}^{3} & -\phi & 0 & -\phi^{4} & 0 & 0 & V^{1} & 0 \\
-\tilde{\phi}^{4} & 0 & -\phi & \phi^{3} & 0 & -V^{1} & 0 & 0\\
0 & -\phi^{4} & \phi^{3} & \bar{\phi} & V^{1} & 0 & 0 & 0 \\
0 & 0 & 0 & V^{1} & -\phi & \phi^{3} & \phi^{4} & 0\\
0 & 0 & -V^{1} & 0 & \phi^{3} & \bar{\phi} & 0 & \phi^{4} \\
0 & V^{1} & 0 & 0 & \phi^{4} & 0 & \bar{\phi} & -\phi^{3} \\
-\tilde{V}_{1} & 0 & 0 & 0 & 0 & \phi^{4} & -\phi^{3} & -\phi \\
\end{array} \right)
\eeq

We cannot decompose the Yukawa matrix as given in the continuum case because of the specific form of the latticization procedure we have chosen.

The action has a residual part also
\beq
S_{R} = \beta \; \textrm{Tr}\; \sum_{x} \psi_{1}^{1}(x)\psi_{1}^{1}(x)D_{1}^{+}\phi^{1}(x)-i\psi_{1}^{1}(x)\psi_{1}^{1}(x)D_{1}^{+}\phi^{2}(x).
\eeq

\section{The Sixteen Supercharge Model}
Starting from the form of the four dimensional 
gauge fermion $\Lambda$ of twisted ${\cal N}=4$ super Yang-Mills theory 
given in \cite{Simon3} we find
\bea
\label{eq:Lambda4}
\Lambda&=& \int d^{4}x ~\textrm{Tr} \Big[~\chi_{\mu \nu}\Big(F_{\mu \nu} +\frac{1}{2}B_{\mu \nu} -\frac{1}{4}[W_{\mu \lambda \rho}, W_{\nu \lambda \rho}]+ \sqrt{2}D_{\lambda}W_{\lambda \mu \nu}\Big)+\psi_{\mu}D_{\mu}\bar{\phi}+\frac{1}{4}\eta[\phi, \bar{\phi}]\nn \\
&&~~~~~~~~~~~~~+\frac{1}{2}\frac{1}{3!}\theta_{\mu \nu \lambda}[W_{\mu \nu \lambda}, \bar{\phi}]-\frac{1}{3!}\kappa_{\mu \nu \lambda \rho} D_{[\mu}W_{\nu \lambda \rho]}+\frac{1}{2}\frac{1}{4!}\kappa_{\mu \nu \lambda \rho}C_{\mu \nu \lambda \rho}\Big].
\eea

To construct a 16 supercharge Yang-Mills quantum mechanics, 
we dimensionally reduce eqn. (\ref{eq:Lambda4}) with 
respect to the $x^{2}$-, $x^{3}$- and $x^{4}$-directions. 
The reduced scalar supercharge transformation rules are
\bea
&&\cQ A_{\mu} = \psi_{\mu},~~~~~~~~~~~~~~~~~~~~~~~~~~\cQ \psi_{1} = - D_{1}\phi, \nn \\
&&\cQ \psi_{i} = -[A_{i}, \phi], \; \; i \neq 1~,~~~~~~~~~~\cQ \phi = 0, \nn \\
&&\cQ \bar{\phi} = \eta,~~~~~~~~~~~~~~~~~~~~~~~~~~~~~~\cQ \eta = [\phi, \bar{\phi}],\nn \\
&&\cQ B_{\mu \nu} = [\phi, \chi_{\mu \nu}],~~~~~~~~~~~~~~~~~~~\cQ \chi_{\mu \nu} = B_{\mu \nu}, \\
&&\cQ W_{\mu \nu \lambda} = \theta_{\mu \nu \lambda},~~~~~~~~~~~~~~~~~~~~~\cQ \theta_{\mu \nu \lambda} = [\phi, W_{\mu \nu \lambda}], \nn \\
&&\cQ C_{\mu \nu \lambda \rho} = [\phi, \kappa_{\mu \nu \lambda \rho}],~~~~~~~~~~~~~~\cQ \kappa_{\mu \nu \lambda \rho} = C_{\mu \nu \lambda \rho}. \nn
\eea 

After $\cQ$-variation and integrating out the multiplier fields we find 
the action
\beq
S = S_{B} + S_{F} + S_{Y}+S_{R}.
\eeq

In terms of the relabelled fields
\bea
&&\psi_{1}^{1} = \psi_{1},~~~~~~~~~~~~~~~~~~\eta^{1} = \eta,~~~~~~~~~~~~~~~V_{1}^{1} = W_{123},\nn \\
&&\psi_{1}^{2} = \chi_{12},~~~~~~~~~~~~~~~~~\eta^{2} = \psi_{2},~~~~~~~~~~~~~V_{1}^{2} = W_{124},\nn \\
&&\psi_{1}^{3} = \chi_{13},~~~~~~~~~~~~~~~~~\eta^{3} = \psi_{3},~~~~~~~~~~~~~V_{1}^{3} = W_{134},\nn \\
&&\psi_{1}^{4} = \theta_{123},~~~~~~~~~~~~~~~~\eta^{4} = \chi_{23},~~~~~~~~~~~~~\phi^{1} = W_{234},\\
&&\psi_{1}^{5} = \chi_{14},~~~~~~~~~~~~~~~~~\eta^{5} = \psi_{4},~~~~~~~~~~~~~~\phi^{2} = A_{2},\nn \\
&&\psi_{1}^{6} = \theta_{124},~~~~~~~~~~~~~~~~\eta^{6} = \chi_{24},~~~~~~~~~~~~~\phi^{3} = A_{3},\nn \\
&&\psi_{1}^{7} = \theta_{134},~~~~~~~~~~~~~~~~\eta^{7} = \chi_{34},~~~~~~~~~~~~~\phi^{4} = A_{4},\nn \\
&&\psi_{1}^{8} = \kappa_{1234},~~~~~~~~~~~~~~\eta^{8} = -\theta_{234},~~~~~~~~~~\phi = \phi^{5} + i \phi^{6},\nn
\eea
the bosonic part of the action takes the form
\bea
S_{B}&=&\beta \; \textrm{Tr}\int dx \Big(-\sum_{i=1}^{6}(D_{1}\phi^{i})^{2} - \sum_{i=1}^{3}(D_{1}V_{1}^{i})^{2} - \sum_{i<j, \; i,j = 1}^{6} \; [\phi^{i}, \phi^{j}]^{2}~~~~~~~~\nn \\
&&~~~~~~~~~~~~~~~~~~~~~~~~~~~~~~~~~~~~~~~~~~~- \sum_{i<j, \; i,j = 1}^{3} \; [V_{1}^{i}, V_{1}^{j}]^{2} - \sum_{i=1}^{3}\sum_{j=1}^{6} \; [V_{1}^{i}, \phi^{j}]^{2}\Big)~,
\eea
where we have made use of integration by parts and the Bianchi identity in
simplifying the original expression. If we were to relabel the fields
$V_{1}^{i},i=1\ldots 3$ as additional scalars this would be the usual
bosonic action of ${\cal N}=1$ super Yang-Mills in $D=10$
reduced to one dimension.

The fermion kinetic term  and Yukawa interactions can then be put
in the  compact form
\beq
S_{F+Y} = \beta \; \textrm{Tr}\; \int dx \; {\bf \Psi}^{\dagger} \; \Gamma^{10}D_{1} {\bf \Psi} + \sum_{i=1}^{6} {\bf \Psi}^{\dagger} [\Gamma^{i}\phi^{i}, {\bf \Psi}]+ \sum_{j=7}^{9} {\bf \Psi}^{\dagger} [\Gamma^{j}V_{1}^{j}, {\bf \Psi}]~,
\eeq
where the spinor
\beq
\label{eq:Spinor16}
{\bf \Psi^{\dagger}} =(\psi_{1}^{1} \; \psi_{1}^{2} \; \psi_{1}^{3} \; \psi_{1}^{4} \; \psi_{1}^{5} \; \psi_{1}^{6} \; \psi_{1}^{7} \; \psi_{1}^{8} \; \eta^{1} \; \eta^{2} \; \eta^{3} \; \eta^{4} \; \eta^{5} \; \eta^{6} \; \eta^{7} \; \eta^{8} \; ),
\eeq
and the $\Gamma$'s,
\bea
&&\Gamma^{1}=-\sigma_{3} \otimes \sigma_{2} \otimes \sigma_{1} \otimes \sigma_{2},~~~\Gamma^{2}=\sigma_{3} \otimes {\bf 1} \otimes {\bf 1} \otimes \sigma_{1},~~~~~~~~\Gamma^{3}=\sigma_{3} \otimes {\bf 1} \otimes \sigma_{1} \otimes \sigma_{3},\nn \\
&&\Gamma^{4}=\sigma_{3} \otimes \sigma_{1} \otimes \sigma_{3} \otimes \sigma_{3},~~~~~\Gamma^{5}=\sigma_{3} \otimes \sigma_{3} \otimes \sigma_{3} \otimes \sigma_{3},~~~~~~\Gamma^{6}=-i {\bf 1} \otimes {\bf 1} \otimes {\bf 1} \otimes {\bf 1},\nn \\
&& \Gamma^{7}=-\sigma_{2} \otimes {\bf 1} \otimes \sigma_{2} \otimes \sigma_{1},~~~~\Gamma^{8}=-\sigma_{2} \otimes \sigma_{2} \otimes \sigma_{3} \otimes \sigma_{1},~~~~\Gamma^{9}=-\sigma_{2} \otimes \sigma_{2} \otimes \sigma_{1} \otimes {\bf 1},\nn \\
&&\Gamma^{10}=\sigma_{1}\otimes {\bf 1} \otimes {\bf 1} \otimes {\bf 1}\nn
\eea

As we
have seen dimensional reduction of the 16 
supercharge action leads to eight scalar fermions
and hence also eight scalar
supercharges among which we have made use of only one. 
Using the scalar supercharge $\cQ$ and the symmetries of the action, 
one can derive the remaining supersymmetries. For example, to 
obtain another scalar super symmetry say $\cQ'$, we look at the invariance of the action under the field transformations

\bea
&& V_{1}^{1} \rightarrow -V_{1}^{1}, ~~~~~~~~\phi^{1} \rightarrow -\phi^{1},\nn \\ 
&&V_{1}^{2} \rightarrow -V_{1}^{2}, ~~~~~~~~\phi^{3} \rightarrow -\phi^{3},
\eea

along with

\bea
&&\psi_{1}^{1} \rightarrow \psi_{1}^{4}, ~~~~~~~~ \psi_{1}^{5} \rightarrow \psi_{1}^{8}, ~~~~~~~~~~~~~~~~ \eta^{1} \rightarrow \eta^{4}, ~~~~~~~~ \eta^{5} \rightarrow \eta^{8}\nn \\
&&\psi_{1}^{2} \rightarrow \psi_{1}^{3}, ~~~~~~~~ \psi_{1}^{6} \rightarrow \psi_{1}^{7}, ~~~~~~~~~~~~~~~~ \eta^{2} \rightarrow \eta^{3}, ~~~~~~~~ \eta^{6} \rightarrow \eta^{7}\nn \\
&&\psi_{1}^{3} \rightarrow \psi_{1}^{2}, ~~~~~~~~ \psi_{1}^{7} \rightarrow \psi_{1}^{6}, ~~~~~~~~~~~~~~~~ \eta^{3} \rightarrow \eta^{2}, ~~~~~~~~ \eta^{7} \rightarrow \eta^{6}\nn \\
&&\psi_{1}^{4} \rightarrow \psi_{1}^{1}, ~~~~~~~~ \psi_{1}^{8} \rightarrow \psi_{1}^{5}, ~~~~~~~~~~~~~~~~ \eta^{4} \rightarrow \eta^{1}, ~~~~~~~~ \eta^{8} \rightarrow \eta^{5}.
\eea

The scalar supersymmetry, $\cQ'$ associated with this invariance of the action is

\bea
&&\cQ' \psi_{1}^{1}= [\phi, V_{1}^{1}],~~~~~~~~~~\cQ' \eta^{1}=B_{23},~~~~~~~~~~~\cQ' B_{12}=-[\phi, \psi_{1}^{3}],~~~~~~~\cQ' \phi^{1}=\eta^{5},\nn \\
&&\cQ' \psi_{1}^{2}= B_{13},~~~~~~~~~~~~~~\cQ' \eta^{2}=-[\phi, \phi^{3}],~~~~~\cQ' B_{13}=[\phi, \psi_{1}^{2}],~~~~~~~~~~\cQ' \phi^{2}= \eta^{3},\nn \\
&&\cQ' \psi_{1}^{3}= B_{12},~~~~~~~~~~~~~~\cQ' \eta^{3}=[\phi, \phi^{2}],~~~~~~~~\cQ' B_{14}=[\phi, \psi_{1}^{8}],~~~~~~~~~\cQ' \phi^{3}= -\eta^{2},\nn \\
&&\cQ' \psi_{1}^{4}= -D_{1}\phi,~~~~~~~~~~~\cQ' \eta^{4}=[\phi, \bar{\phi}],~~~~~~~~~\cQ' B_{23}=[\phi, \eta^{1}],~~~~~~~~~~\cQ' \phi^{4}= \eta^{8},\nn \\
&&\cQ' \psi_{1}^{5}=C_{1234},~~~~~~~~~~~~\cQ' \eta^{5}=[\phi, \phi^{1}],~~~~~~~~\cQ' B_{24}=[\phi, \eta^{7}],~~~~~~~~~~\cQ' V_{1}^{1}=-\psi_{1}^{1}, \\
&&\cQ' \psi_{1}^{6}=[\phi, V_{1}^{3}],~~~~~~~~~~~\cQ' \eta^{6}=B_{34},~~~~~~~~~~~\cQ' B_{34}=[\phi, \eta^{6}],~~~~~~~~~~\cQ' V_{1}^{2}=\psi_{1}^{7},\nn \\
&&\cQ' \psi_{1}^{7}=-[\phi, V_{1}^{2}],~~~~~~~~\cQ' \eta^{7}=B_{24},~~~~~~~~~~~~\cQ' A^{1}= \psi_{1}^{4},  ~~~~~~~~~~~~~~~\cQ' V_{1}^{3}=\psi_{1}^{6},\nn \\
&&\cQ' \psi_{1}^{8}=B_{14},~~~~~~~~~~~~~~\cQ' \eta^{8}=[\phi, \phi^{4}],~~~~~~~~~\cQ' \phi=0,~~~~~~~~~~~~~~~~~~~\cQ' \bar{\phi} = \eta^{4},\nn \\
&&\cQ' C_{1234}=[\phi, \psi_{1}^{5}].\nn
\eea
{\ }\\
In principle Ward identities corresponding to these additional scalar
supersymmetries can be computed in lattice Monte Carlo simulations to test
for a restoration of full supersymmetry in the continuum limit.

Finally, to construct the lattice version of the theory, 
we again write down the dimensionally reduced $\Lambda$ on a lattice. 
The $\cQ$-transformation rules for fields on lattice are similar to
the $\cQ=4$ and $\cQ=8$ models

\bea
&&\cQ U_{1}(x) = \psi_{1}(x)U_{1}(x)~~~~~~~~~~~~~~~~~~~~~~~~\cQ A_{\mu}(x) = \psi_{\mu}(x)\nn \\
&&\cQ \psi_{1}(x) = \psi_{1}(x)\psi_{1}(x) - D_{1}^{+}\phi(x),~~~~~~~~~\cQ \psi_{i}(x) = -[A_{i}(x), \phi(x)], \; \; i \neq 1,\nn \\
&&\cQ \eta(x) = [\phi(x), \bar{\phi}(x)],~~~~~~~~~~~~~~~~~~~~~~~~~\cQ B_{\mu \nu}(x) = [\phi(x), \chi_{\mu \nu}(x)]\nn \\
&&\cQ \chi_{\mu \nu}(x) = B_{\mu \nu}(x),~~~~~~~~~~~~~~~~~~~~~~~~~~~\cQ W_{\mu \nu \lambda}(x) = \theta_{\mu \nu \lambda}(x) \\
&&\cQ \theta_{\mu \nu \lambda}(x) = [\phi(x), W_{\mu \nu \lambda}(x)],~~~~~~~~~~~~~~~~\cQ C_{\mu \nu \lambda \rho}(x) = [\phi(x), \kappa_{\mu \nu \lambda \rho}(x)]\nn \\
&&\cQ \kappa_{\mu \nu \lambda \rho}(x) = C_{\mu \nu \lambda \rho}(x),~~~~~~~~~~~~~~~~~~~~~~~\cQ \phi(x) = 0,\nn \\
&&\cQ \bar{\phi}(x) = \eta(x).\nn
\eea 

Carrying out the $\cQ$-variation we write down the action

\beq
S = S_{B} + S_{F} + S_{Y}+S_{R}.
\eeq

In terms of the relabelled fields

\bea
&&\psi_{1}^{1}(x) = \psi_{1}(x),~~~~~~~~~~~~~\eta^{1}(x) = \eta(x),~~~~~~~~~~~~~~~~V_{1}^{1}(x) = W_{123}(x),~~~~~~\nn \\
&&\psi_{1}^{2}(x) = \chi_{12}(x),~~~~~~~~~~~~\eta^{2}(x) = \psi_{2}(x),~~~~~~~~~~~~~~V_{1}^{2}(x) = W_{124}(x),~~~~~~\nn \\
&&\psi_{1}^{3}(x) = \chi_{13}(x),~~~~~~~~~~~~\eta^{3}(x) = \psi_{3}(x),~~~~~~~~~~~~~~V_{1}^{3}(x) = W_{134}(x),~~~~~~\nn \\
&&\psi_{1}^{4}(x) = \theta_{123}(x),~~~~~~~~~~~\eta^{4}(x) = \chi_{23}(x),~~~~~~~~~~~~~~\phi^{1}(x) = W_{234}(x),~~~~~~\\
&&\psi_{1}^{5}(x) = \chi_{14}(x),~~~~~~~~~~~~\eta^{5}(x) = \psi_{4}(x),~~~~~~~~~~~~~~~\phi^{2}(x) = A_{2}(x),~~~~~~\nn \\
&&\psi_{1}^{6}(x) = \theta_{124}(x),~~~~~~~~~~~\eta^{6}(x) = \chi_{24}(x),~~~~~~~~~~~~~~\phi^{3}(x) = A_{3}(x),~~~~~~\nn \\
&&\psi_{1}^{7}(x) = \theta_{134}(x),~~~~~~~~~~~\eta^{7}(x) = \chi_{34}(x),~~~~~~~~~~~~~~\phi^{4}(x) = A_{4}(x),~~~~~~\nn \\
&&\psi_{1}^{8}(x) = \kappa_{1234}(x),~~~~~~~~~~\eta^{8}(x) = -\theta_{234}(x),~~~~~~~~~~~\phi(x) = \phi^{5}(x) + i \phi^{6}(x),~~~~~~\nn
\eea

the bosonic part takes the form

\bea
S_{B} &=& \beta \int dx \; \textrm{Tr} \; \Big[ -\Big(D^+_{1}\phi^{2}-
[\phi^{1}, V_{1}^{3}]+[\phi^{3},V_{1}^{1}]-
[\phi^{4},V_{1}^{2}] \Big)^{2}- \Big(D^+_{1}\phi^{3}-
[\phi^{1}, V_{1}^{2}]+[\phi^{4},V_{1}^{3}]\nn \\
&&~~~-[\phi^{2},V_{1}^{1}] \Big)^{2} - \Big(D^+_{1}\phi^{4}-[\phi^{1}, V_{1}^{1}]+
[\phi^{2},V_{1}^{2}]-[\phi^{3},V_{1}^{3}] \Big)^{2}-\Big([\phi^{2}, \phi^{3}]-
[V_{1}^{3}, V_{1}^{2}]\nn \\
&&~~~+D^{-}_{1}V_{1}^{1}-[\phi^{4},\phi^{1}] \Big)^{2}-\Big([\phi^{2}, \phi^{4}]-[V_{1}^{3}, V_{1}^{1}]+[\phi^{3},\phi^{1}] -D^{-}_{1}V_{1}^{2} \Big)^{2}- \Big([\phi^{3},\phi^{4}]\nn \\
&&~~~-[V_{1}^{2}, V_{1}^{1}]+
D^{-}_{1}V_{1}^{3}-[\phi^{2}, \phi^{1}] \Big)^{2}+\frac{1}{2}\Big(D^+_{1} \phi^{1}\Big)^{2}+\frac{1}{2}[\phi^{2}, V_{1}^{3}]^{2}+\frac{1}{2}[\phi^{3}, V_{1}^{2}]^{2}\nn \\
&&~~~+\frac{1}{2}[\phi^{4}, V_{1}^{1}]^{2}-
\Big( (D^+_{1}\phi^{5})^{2}+(D^+_{1}\phi^{6})^{2}+[\phi^{2}, \phi^{5}]^{2}+
[\phi^{2}, \phi^{6}]^{2}+[\phi^{3}, \phi^{5}]^{2}\nn \\
&&~~~+[\phi^{3}, \phi^{6}]^{2}+[\phi^{4}, \phi^{5}]^{2}+
[\phi^{4}, \phi^{6}]^{2}\Big)- [\phi^{5}, \phi^{6}]^{2}-
\Big([\phi^{5}, \phi^{1}]^{2}+[\phi^{6}, \phi^{1}]^{2}+
[\phi^{5}, V_{1}^{3}]^{2}\nn \\
&&~~~+[\phi^{6}, V_{1}^{3}]^{2}+[\phi^{5}, V_{1}^{2}]^{2}+
[\phi^{6}, V_{1}^{2}]^{2}+[\phi^{5}, V_{1}^{1}]^{2}+
[\phi^{6}, V_{1}^{1}]^{2}\Big)\Big]~.
\eea
{\ }\\
Notice that here also this cannot be further simplified
since the lattice theory does not possess an exact Bianchi identity.
The fermionic kinetic term can be expressed in a matrix form similar to the $\Gamma^{10}$ matrix in the continuum, with the same spinor structure as eqn. (\ref{eq:Spinor16}),

\beq
{\bf D}_{1} = \left( \begin{array}{cc}
0 &  ({\mathbb I}_{8 \times 8})D_{1}^{-} \\
({\mathbb I}_{8 \times 8})D_{1}^{+} & 0 \end{array} \right)
\eeq
{\ }\\
The Yukawa part of the action $S_{Y}$ can also be expressed in a matrix form

\beq
{\bf \Psi^{\dagger}[\tilde{O}, \Psi]},
\eeq
where ${\bf \tilde{O}}$ is a $16 \times 16$ matrix: 

\begin{center}
$\left( \begin{array}{cccccccccccccccc}
\tilde{\bar{\phi}} & \tilde{\phi}^{2} & \tilde{\phi}^{3} & 0 & \tilde{\phi}^{4} & 0 & 0 & \tilde{\phi}^{1} & 0 & 0 & 0 & \tilde{V}_{1}^{1} & 0 & \tilde{V}_{1}^{2} & \tilde{V}_{1}^{3} & 0 \\
\tilde{\phi}^{2} & -\phi & 0 & -\phi^{3} & 0 & -\phi^{4} & -\phi^{1} & 0 & 0 & 0 & V_{1}^{1} & 0 & V_{1}^{2} & 0 & 0 & V_{1}^{3} \\
\tilde{\phi}^{3} & 0 & -\phi & \phi^{2} & 0 & \phi^{1} & -\phi^{4} & 0 & 0 & -V_{1}^{1} & 0 & 0 & V_{1}^{3} & 0 & 0 & -V_{1}^{2} \\
0 & -\phi^{3} & \phi^{2} & \bar{\phi} & -\phi^{1} & 0 & 0 & \phi^{4} & -V_{1}^{1} & 0 & 0 & 0 & 0 & V_{1}^{3} & -V_{1}^{2} & 0 \\
\tilde{\phi}^{4} & 0 & 0 & -\phi^{1} & -\phi & \phi^{2} & \phi^{3} & 0 & 0 & -V_{1}^{2} & -V_{1}^{3} & 0 & 0 & 0 & 0 & V_{1}^{1} \\
0 & -\phi^{4} & \phi^{1} & 0 & \phi^{2} & \bar{\phi} & 0 & -\phi^{3} & -V_{1}^{2} & 0 & 0 & -V_{1}^{3} & 0 & 0 & V_{1}^{1} & 0 \\
0 & -\phi^{1} & -\phi^{4} & 0 & \phi^{3} & 0 & \bar{\phi} & \phi^{2} & -V_{1}^{3} & 0 & 0 & V_{1}^{2} & 0 & -V_{1}^{1} & 0 & 0 \\
\tilde{\phi}^{1} & 0 & 0 & \phi^{4} & 0 & -\phi^{3} & \phi^{2} & -\phi & 0 & -V_{1}^{3} & V_{1}^{2} & 0 & -V_{1}^{1} & 0 & 0 & 0 \\
0 & 0 & 0 & -V_{1}^{1} & 0 & -V_{1}^{2} & -V_{1}^{3} & 0 & -\phi & -\phi^{2} & -\phi^{3} & 0 & -\phi^{4} & 0 & 0 & -\phi^{1} \\
0 & 0 & -V_{1}^{1} & 0 & -V_{1}^{2} & 0 & 0 & -V_{1}^{3} & -\phi^{2} & \bar{\phi} & 0 & \phi^{3} & 0 & \phi^{4} & \phi^{1} & 0 \\
0 & V_{1}^{1} & 0 & 0 & -V_{1}^{3} & 0 & 0 & V_{1}^{2} & -\phi^{3} & 0 & \bar{\phi} & -\phi^{2} & 0 & -\phi^{1} & \phi^{4} & 0 \\
\tilde{V}_{1}^{1} & 0 & 0 & 0 & 0 & -V_{1}^{3} & V_{1}^{2} & 0 & 0 & \phi^{3} & -\phi^{2} & -\phi & \phi^{1} & 0 & 0 & -\phi^{4} \\
0 & V_{1}^{2} & V_{1}^{3} & 0 & 0 & 0 & 0 & -V_{1}^{1} & -\phi^{4} & 0 & 0 & \phi^{1} & \bar{\phi} & -\phi^{2} & -\phi^{3} & 0 \\
\tilde{V}_{1}^{2} & 0 & 0 & V_{1}^{3} & 0 & 0 & -V_{1}^{1} & 0 & 0 & \phi^{4} & -\phi^{1} & 0 & -\phi^{2} & -\phi & 0 & \phi^{3} \\
\tilde{V}_{1}^{3} & 0 & 0 & -V_{1}^{2} & 0 & V_{1}^{1} & 0 & 0 & 0 & \phi^{1} & \phi^{4} & 0 & -\phi^{3} & 0 & -\phi & -\phi^{2} \\
0 & V_{1}^{3} & -V_{1}^{2} & 0 & V_{1}^{1} & 0 & 0 & 0 & -\phi^{1} & 0 & 0 & -\phi^{4} & 0 & \phi^{3} & -\phi^{2} & \bar{\phi} \\
\end{array} \right)$
\end{center}
{\ }\\
Notice that here also we cannot decompose the Yukawa matrix as given in the continuum case because of the specific form of the latticization procedure we have chosen.

Here also the lattice action picks up a residual part
\beq
S_{R} = \beta \; \textrm{Tr}\; \sum_{x} \psi_{1}^{1}(x)\psi_{1}^{1}(x)D_{1}^{+}\phi_{1}(x)-i\psi_{1}^{1}(x)\psi_{1}^{1}(x)D_{1}^{+}\phi_{2}(x).
\eeq
\section{Conclusions}
In this paper we have derived twisted actions for $\cQ=4, 8$ and $16$
supercharge Yang-Mills quantum mechanics by dimensionally reducing
the corresponding $\cQ$-exact actions in $D=2, 3$ and $4$ dimensions. 
The dimensional reduction is done in the continuum but the final theories
may be translated to the lattice using a simple discretization prescription
that preserves the $\cQ$-supersymmetry. We show that the reduced twisted
theories 
contain two, four or eight one dimensional K\"{a}hler-Dirac fields respectively.
Correspondingly the continuum models contain additional nilpotent scalar
supercharges all but one of which are softly 
broken by the descretization procedure adopted here. These lattice theories should prove useful in studies of
the holographic correspondence between maximally supersymmetric
Yang-Mills quantum mechanics and D0-branes in type IIA string theory.
It would also be interesting to compare these lattice actions with those
derived by orbifold methods as detailed in \cite{Kaplan,orbifold}.
\section{Acknowledgements}
SC would like to thank Toby Wiseman for useful discussions. This work is supported in part by the US Department of Energy grant DE-FG02-85ER40237. 

\bibliographystyle{apsrmp}

\begin{thebibliography}{99}
\bibitem{Comment1}{Although it appears that the breaking of SUSY in \cite{Jun,Nishimura} is rather small}
\end{thebibliography}

\end{document}